\documentclass[12pt]{article}
\usepackage{amsmath}
\usepackage[dvips]{graphicx}
\usepackage{epsfig}
\usepackage{amsmath}
\usepackage{amssymb}
\usepackage{graphics,epstopdf}
\usepackage{amsfonts}
\usepackage{amsthm}
\usepackage[usenames]{color}

\definecolor{AV}{rgb}{0.65,0.0,0}
\definecolor{GC}{rgb}{0,0.0,0.65}
\definecolor{WS}{rgb}{0,0.65,0}

\usepackage[
      colorlinks=true,
      linkcolor=blue,
      urlcolor=blue,
      filecolor=blue,
      citecolor=red,
      pdfstartview=FitV,
      pdftitle={},
      pdfauthor={},
      pdfsubject={},
      pdfkeywords={},
      pdfpagemode=None,
      bookmarksopen=true
]{hyperref}

\newcommand{\bm}{\begin{multiline}}
\newcommand{\beq}{\begin{equation}}
\newcommand{\eeq}{\end{equation}}
\newcommand{\beqs}{\begin{eqnarray}}
\newcommand{\eeqs}{\end{eqnarray}}

\begin{document}

\title{\textbf{Charged particles moving around a spherically symmetric
dilatonic black hole}}

\author{\textbf{\textbf{Vitalie Lungu}$^1$\footnote{E-mail: vitalie.lungu@student.uaic.ro},  Marina--Aura Dariescu}$^2$\footnote{E-mail: marina@uaic.ro}, \textbf{Ciprian Dariescu}$^3$\footnote{E-mail: ciprian.dariescu@uaic.ro} \\  \textbf{and Cristian Stelea}$^{4}$\footnote{Corresponding author e-mail: \texttt{cristian.stelea@uaic.ro}} \\
$^{1,2,3}$Faculty of Physics, 
{ ``Alexandru Ioan Cuza"} University  of Iasi \\
Bd. Carol I, No. 11, 700506 Iasi, Romania \\
$^{4}$Department of Exact and Natural Sciences, \\
Institute of Interdisciplinary Research,\\
 { ``Alexandru Ioan Cuza"} University of Iasi, \\
 Bd. Carol I, no. 11, 700506 Iasi, Romania 
}

\date{}
\maketitle

\begin{abstract}
For the static spherically symmetric dilatonic black hole described by the Gibbons–Maeda–Garfinkle–Horowitz–Strominger geometry, we analyze the timelike trajectories for electrically charged test particles. Both cases of an electric black hole and a magnetic one are considered.
Finally, we are obtaining the solution to the Klein--Gordon equation in terms of Heun confluent functions and the corresponding energy spectrum. A special attention is given to the role of the dilaton parameter.
\end{abstract}

\begin{flushleft}
{\it Keywords}: Black Holes; Klein--Gordon equation; Heun functions.
 \\
{\it PACS:}
04.20.Jb Exact solutions;
02.40.Ky Riemannian geometries;
04.62.+v Quantum fields in curved spacetimes;
02.30. Gp Special functions.
\end{flushleft}

\baselineskip 1.5em

\newpage

\section{Introduction}

In the last years, different types of dilatonic black hole (BH) solutions have been used for testing theories of gravity by means of black hole shadows \cite{Mizuno:2018lxz}. Working in the context of low-energy heterotic string theory compactified to four dimensions, a spherically symmetric static solution describing a charged black hole in presence of a scalar field was found by Garfinkle, Horowitz and Strominger \cite{Garfinkle:1990qj}, while the same solution was initially derived by Gibbons and Maeda in 1988 \cite{GM}. This metric is now known as the GMGHS solution and, in the last thirty years, it has received considerable attention from the physical community.

Besides the BH's gravitational mass $M$ and electromagnetic charge $Q$, the general solution contains the asymptotic value of the scalar dilaton field, $\phi_0$, while the scalar field has a dilatonic `secondary hair' $D\propto \frac{Q^2}{M}e^{\phi_0}$ \cite{Garfinkle:1990qj}, which is also conserved and this has important physical consequences. In our work we shall consider for simplicity the asymptotic value of the dilaton to be $\phi_0=0$, such that the dilaton scalar charge is proportional to the dilaton parameter $r_0=\frac{Q^2}{M}$.

The rotating version of the GMGHS solution was found by Sen \cite{Sen:1992ua} in context of the so-called EMDA theory. The EMDA theory, which is obtained from the compactification of the ten-dimensional heterotic string theory on a six-dimensional torus, contains besides the metric a dilaton, a Maxwell field and a pseudoscalar axionic field. This rotating solution is characterized by the mass parameter $M$, the electric charge $Q$ and the angular momentum per unit mass, $a$. In absence of rotation, when $a=0$, it reduces to the static spherically symmetric GMGHS solution, with the dilatonic parameter $r_0=\frac{Q^2}{M}$. On the other hand, if the electric charge vanishes then the Kerr-Sen solution reduces to the vacuum Kerr solution. 
As a low-energy effective field theory of the heterotic string theory, the EMDA theory attracted a lot of attention in recent years. It is valuable to investigate the role of such a theory in astrophysical observations, as an alternative to the usual General Relativity \cite{Narang:2020bgo}.

Recently, the range of the dilaton parameter $r_0=\frac{Q^2}{M}$ has been estimated by monitoring
the geodesic motions of stars in our Galactic Center. Among the short-period stars, the S2 star with a 16 year orbit around Sagittarius A* (Sgr A*) has been seen as a very attractive target. In 2020, the GRAVITY Collaboration reported the first detection of the General Relativity Schwarzschild precession in its orbit \cite{GRAVITY:2020gka}. Using these results, in \cite{Fernandez:2023kro} the dilaton parameter was constrained to $r_0\lesssim 0.066M$.  In \cite{Banerjee:2020qmi}, a preferred value of $r_0\thickapprox 0.2M$ was determined using the optical continuum spectrum of quasars. Finally, by observations of the shadow diameters of M87* and Sqr A*, in \cite{Sahoo:2023czj} the dilaton parameter was also constrained to the interval $0.1M\lesssim r_0\lesssim 0.4M$. More recently, in \cite{Tripathi:2021rwb} a more stringent constraint $r_0<0.011M$ was found using X-ray reflection spectroscopy observations. Further studies of the EMDA theory include  investigation of the accretion processes in black holes \cite{Banerjee:2020ubc} - \cite{Feng:2024iqj}, the use of a Kerr-Sen black hole as a particle accelerator for spinning particles \cite{An:2017hby}, studies of exact solutions of the Klein-Gordon equation in the Kerr-Sen background \cite{Wu:2001xh}, \cite{Vieira:2018hij}, as well as the connection between the quasinormal modes (QNMs) to the black hole shadow of a rotating Kerr-Sen black hole \cite{Wu:2021pgf}, \cite{Cardoso:2008bp}.

From a theoretical point of view, the null and timelike trajectories of uncharged particles around the GMGHS black hole have been extensively worked out. There is a rich literature dealing with this subject, see for example \cite{Blaga:2014spa} - \cite{Soroushfar:2016yea}. 
However, the case of the charged particles to which the first part of the present work is dedicated is much more complicated.

In our work we made use of a Lagrangean approach to derive the corresponding equations of motion in both the electrically charged case as well as for the magnetically charged black hole. In the purely electric potential, our results can be compared to those derived by Villanueva and Olivares, who solved the equations of motion using the Hamilton--Jacobi method \cite{Villanueva:2015kua}. In this context, one would expect an excessively large rate of orbital precession due to the stronger attraction between the test particle and the central source. 

In particular, our results in the magnetic case can be compared to those obtained for instance in \cite{Gonzalez:2017kxt}. While the full equations of motion for charged test particles in the magnetic case can be integrated exactly, using our Lagrangean approach one can easily show that the motion of electrically charged particles in presence of a magnetically charged black hole is confined to the so-called Poincar\'e cones of various angles. In particular, using our method it is easy to determine the characteristics of the Poincar\'e cones on which the motion of electrically charged particles is bounded to and their dependence on the physical quantities describing the GMGHS geometry and the charged test particle. In particular, we also investigate the existence of circular orbits located at the intersection of a Poincar\'e cone with a sphere.

On the other hand, it is well known that perturbations of a black hole spacetime can
be investigated by considering relativistic particles
evolving in the corresponding manifold. That is why in the second part of the paper, we discuss the Klein--Gordon equation for charged particles moving in the background of a GMGHS black hole. Again, for uncharged particles, the Gordon equation has been investigated in \cite{Mondal:2020pop} and a relation between null geodesics and quasinormal modes frequency in the eikonal approximation has been found. 

In the case considered in the present paper, the corresponding Klein--Gordon equation written for charged particles contains additional terms due to the interaction with the electric or magnetic fields. The solution to the radial equation is obtained in terms of Heun confluent functions \cite{Heun} - \cite{Vieira:2016ubt}.
The so-called resonant frequencies are essential characteristics of a black hole and they can be derived by imposing the requirement that  the Heun function can be cast into a polynomial form \cite{Vieira:2016ubt}.

In general, the Heun functions are unique local Frobenius solutions to a second-order linear ordinary differential equation of the Fuchsian type with $4$ regular singular points. Once the singularities coalesce, one can obtain the so-called confluent Heun functions with two regular and one irregular singularities. The main advantage is that these functions provide analytical solutions to a high number of problems encountered in theoretical and applied sciences. That is why, in the last two decades, there is a raising number of articles on Heun general or confluent functions in view of their wide range of applications. However, especially because of their singularities, there are unsolved problems related to their normalization, series expansions and integration techniques. That is why, many specialists are still relying on approximate methods.

In relation to exact solutions of the Klein-Gordon equation, the theory of the Heun functions provides a direct way of obtaining the quasinormal frequencies (QNMs) by imposing that the Heun functions get a polynomial form. Next, one may use this result to compute the temperature on the black hole’s horizon and the corresponding emission rate. The obtained results agree with those derived using the standard WKB method \cite{Cho:2003qe}.

The present paper is organized as it follows. In the next section we introduce the GMGHS geometry. Note that this geometry can be sourced by an electromagnetic field using either an electric ansatz (in which case the black hole is electrically charged) or a magnetic ansatz (in which case the black hole carries a magnetic charge). The motion of electrically charged test particles will have different characteristics in each case. In Section $2.1$ we consider the electrically charged black hole and study the effective potential and timelike trajectories of charged particles in this background. In Section $2.2$ we consider the case of a magnetically charged black hole. In this case the motion of a charged particle will be confined on Poincar\'e cones of various angles. We also address in Section $2.3$ the case of circular motions on the Poincar\'e cones. They correspond to trajectories of constant $r$ which are the intersection of a sphere with the Poincar\'e cone.
In Section $3$ we are dealing with the Klein--Gordon equation and its analytical solution. A special attention is given to the energy spectrum and its dependence on the dilatonic parameter. The last section is dedicated to conclusions and avenues for further work.

\section{Charged particles in the GMGHS geometry}

\subsection{The GMGHS geometry}

Let us start with the static spherically symmetric charged dilatonic black hole line element known as the Gibbons–Maeda–Garfinkle–Horowitz–Strominger (GMGHS) solution:
\begin{equation}
ds^2= - f(r) dt^2  +  f(r)^{-1}  dr^2 + p^2 \left[  d \theta ^2 + \sin^2 \theta d \varphi^2 \right],
\label{metric}
\end{equation}
where
\begin{equation}
f(r) = 1 - \frac{2M}{r},  
\end{equation}
and 
\begin{equation}
p^2 = r(r-r_0) \; , \quad r_0 = \frac{Q^2}{M}
\end{equation}
In the above expressions, $M$ and $Q$ are the mass and charge of the black hole and
the dilaton field is given by:
\[
e^{2  \phi} = 1 - \frac{Q^2}{Mr}.
\]
This geometry is sourced by a dilaton field $\phi$, which is coupled to the electromagnetic field in the action. Note that, for simplicity, we have chosen in the above expressions the asymptotic value of the dilaton field as $\phi_0=0$ \cite{Garfinkle:1990qj}. One can use the magnetic ansatz for the electromagnetic potential $A_{\mu}=(0, 0, 0, -Q_m\cos\theta)$, in which case the black hole carries a magnetic charge. In the electrically charged case one has $A_{\mu}=(-Q/r, 0, 0, 0)$ and the black hole is now endowed with an electric charge. In both cases the black hole geometry remains the same as in (\ref{metric}), although in the electric case the dilaton field changes its sign.

This charged black hole has a regular event horizon at $r_h = 2M$, which is identical to
the Schwarzschild one, while the theoretical range of the parameter $r_0$ is $0 < r_0 < 2M$.
However, in the last years, there were efforts to put an experimental bound on $r_0$ using astrophysical observations \cite{Banerjee:2020qmi}. Recently, within a fully relativistic orbital model for the S2 star in the Galactic Center of the Milky Way,
it was shown that improved astrometric precision can narrow down the dilaton parameter.
Moreover, by taking into account the information about the orbital precession, the upper limit on $r_0$ has been set at
$r_0 \leq 1.6 M$ \cite{Fernandez:2023kro}.

If one considers now a charged particle moving in the electric field generated by the electromagnetic potential $A_{\mu}$ the motion will be described by the Lagrangean:
\beqs
{\cal L}=\frac{1}{2}g_{\mu\nu}\dot{x}^{\mu}\dot{x}^{\nu}+q A_{\mu}\dot{x}^{\mu},
\eeqs
where $q=e/ m$ is the specific charge of the test particle with charge $e$ and mass $m$.

\subsection{The electrically charged black hole}

In the followings, for the line element (\ref{metric}) we shall analyze how the value of the black hole's charge is affecting the charged particles trajectories.
We focus first on the case of an electric black hole, with the electromagnetic tensor $F_{\mu\nu}$ generated by the four-vector potential 
\begin{equation}
A_t = - \, \frac{Q}{r} \,  ,
\label{At}
\end{equation}
while the magnetic ansatz will be discussed later.

Compared to the motion of uncharged particles \cite{Blaga:2014spa} - \cite{Fernando:2011ki}, in
the case of the particle of charge $q$ and unit mass moving along timelike geodesics, the situation is more complicated, since the Lagrangean contains an additional coupling term, 
\begin{equation}
{\cal L} = \frac{1}{2} \left[ - f \dot{t}^2 + \frac{\dot{r}^2}{f} + p^2 ( \dot{\theta}^2+ \sin^2 \theta \dot{\varphi}^2 ) \right]  - \frac{q Q \dot{t}}{r}.
\end{equation}

For the cyclic coordinates $t$ and $\varphi$, one can define two conserved quantities, i.e. the energy
\begin{equation}
E = f \dot{t} + \frac{q Q}{r}
\end{equation}
and the angular momentum
\begin{eqnarray}
& &
L = p^2 \sin^2 \theta \dot{\varphi}. 
\end{eqnarray}
Following the usual procedure, we replace them in  the normalization condition $g_{\mu\nu}\dot{x}^{\mu}\dot{x}^{\nu}=-1$:
\begin{equation}
 \frac{\dot{r}^2}{f} + p^2 ( \dot{\theta}^2+ \sin^2 \theta \dot{\varphi}^2 )    - f \dot{t}^2 = - 1,
\label{tau}
\end{equation}
where $dot$ are the derivatives with respect to the proper time $\tau$. Similarly to the Schwarzschild case, the geodesics are planar and therefore, one may consider the particles moving in the equatorial plane ($\theta = \pi /2$). One finds the important relation
\begin{equation}
\dot{r}^2 = \left[ E - \frac{qQ}{r} \right]^2  - f \left[ 1 + \frac{L^2}{p^2} \right]  = \left[ E - V_+ \right] \left[ E- V_- \right],
\label{radialel}
\end{equation}
where
\begin{equation}
V_{\pm} \, = \, \frac{q Q}{r} \pm\sqrt{f  \left( 1 + \frac{L^2}{p^2} \right) }.
\label{potel}
\end{equation}
On the event horizon $r_h =2M$, the above expressions for the electric potential (\ref{potel}) become $V_h =q Q/(2M)$ and can be either positive or negative, depending on the sign of $q Q$. For large values of the radial coordinate these potentials tend to $1$.

Since the negative branch $V_-$ has no classical interpretation being associated with antiparticles in the framework
of quantum field theory, from now on, we shall consider the positive potential $V_+$ given in (\ref{potel}) as being the effective potential in which the particle is evolving.
The first term in $V_+$ represents the Coulomb interaction of the particle of charge $q$ with the charged black hole, while the second term corresponds to the neutral particle case.
Depending on the sign of $q Q$, the Coulomb contribution in the potential can be either attractive or repulsive.

In terms of rescaled quantities:
\[
x = \frac{r}{2M} \; , \quad \tilde{Q} = \frac{Q}{2M} \; , \quad \tilde{L} = \frac{L}{2M} \; , \quad a = \frac{r_0}{4M} = \tilde{Q}^2,
\]
the effective potential reads:
\begin{equation}
V_{eff} =  \frac{q \tilde{Q}}{x} +  \sqrt{\left( 1 - \frac{1}{x} \right) \left[ 1 + \frac{{\tilde{L}}^2 }{x(x-2a)} \right] }.
\label{poteleff}
\end{equation}

\begin{figure}
  \centering
  \includegraphics[width=0.7\textwidth]{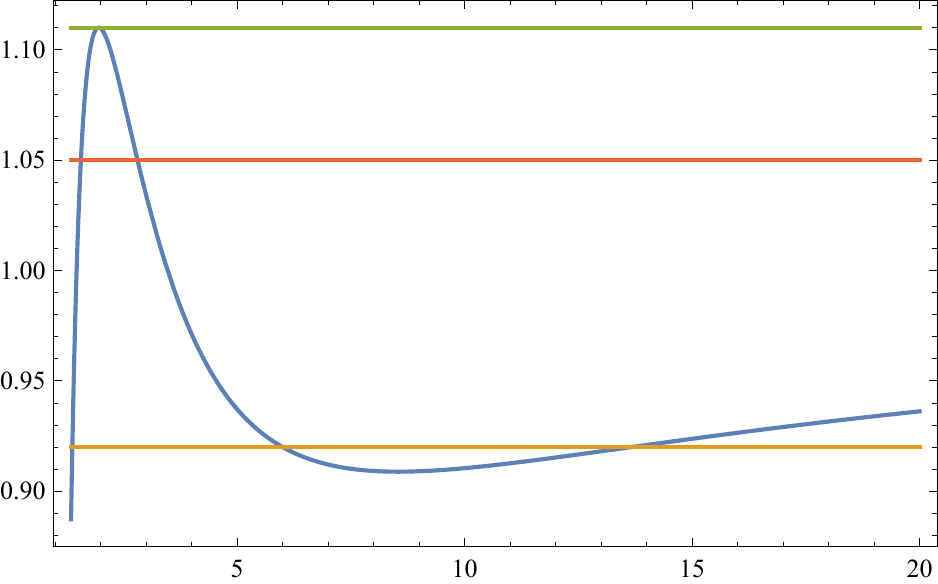}
  \caption{The effective potential given in (\ref{poteleff}), against the rescaled radial coordinate $x = r/(2M)$. The numerical values are: $\tilde{Q} = 0.3$, $q=-4$, $\tilde{L} =4.2$. The horizontal yellow, red and 
green lines correspond to the particle's energy and $V_{max} =1.11$.}
\label{fig1}
\end{figure}

The potential (\ref{poteleff}) is represented in Figure \ref{fig1} and one may notice that it allows different types of trajectories, for the specified values of the black hole's charge and particle's energy and angular momentum. Thus, for $1<E<V_{max}$ the particle whose energy is represented by the red horizontal line has unbounded trajectories. There are two turning points, denoted by $x_1$ and $x_2$ (from left to right), solutions of the equation $E=V_{eff}$. Thus, the particle starting from  $x_0 > x_2$ reaches the turning point $x_2$ and goes back to infinity, while the particle coming from $1<x_0 <x_1$ will be attracted into the black hole. 

\begin{figure}
  \centering
  \includegraphics[width=0.45\textwidth]{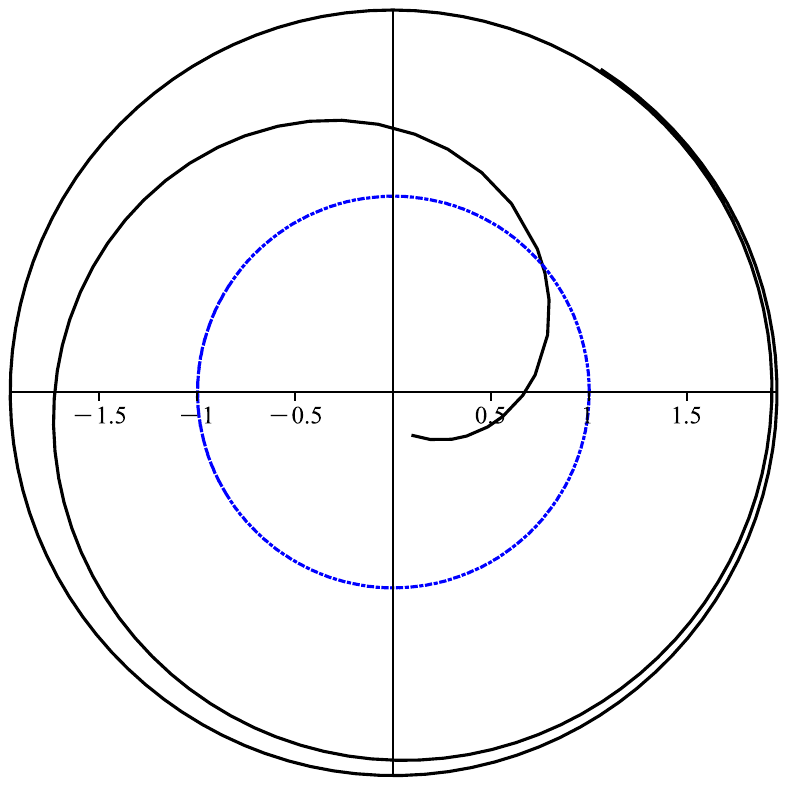} \hspace{0.2cm}
\includegraphics[width=0.45\textwidth]{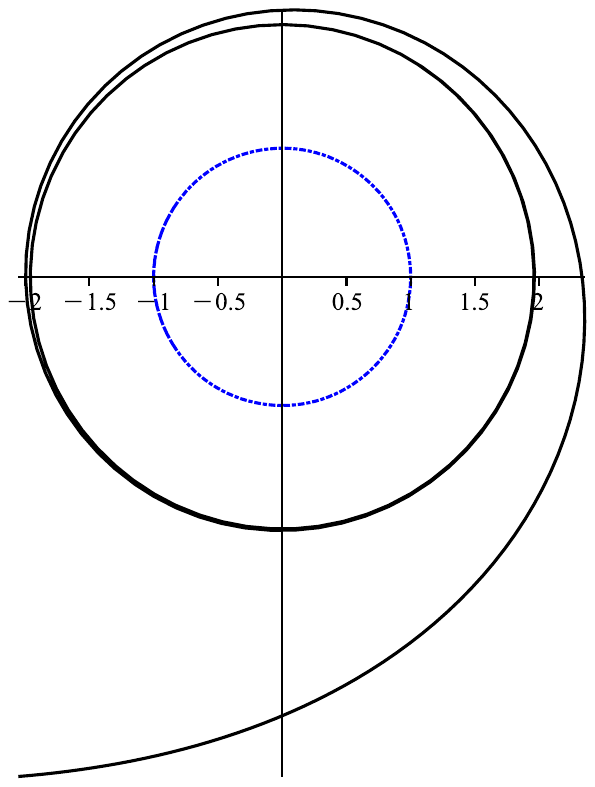} 
  \caption{{\it Left panel.} Unstable circular orbit for the particle with $E=V_{max}= 1.11$ represented by the green horizontal line in Figure \ref{fig1}.
The initial coordinate is $x_0 = 1.9580$. {\it Right panel.} Unstable circular orbit for the particle with $E=V_{max}= 1.11$ and the starting point $x_0 = 1.9583$. The blue circle is the horizon.} \label{fig2}
\end{figure}

The particle with $E= V_{max}$ (the green horizontal line) has an unstable circular orbit.
Depending on the starting point $x_0$, the particle will eventually tend to the singularity (the left panel in Figure \ref{fig2}) or goes to infinity (the right panel in Figure \ref{fig2}).

\begin{figure}
  \centering
  \includegraphics[width=0.45\textwidth]{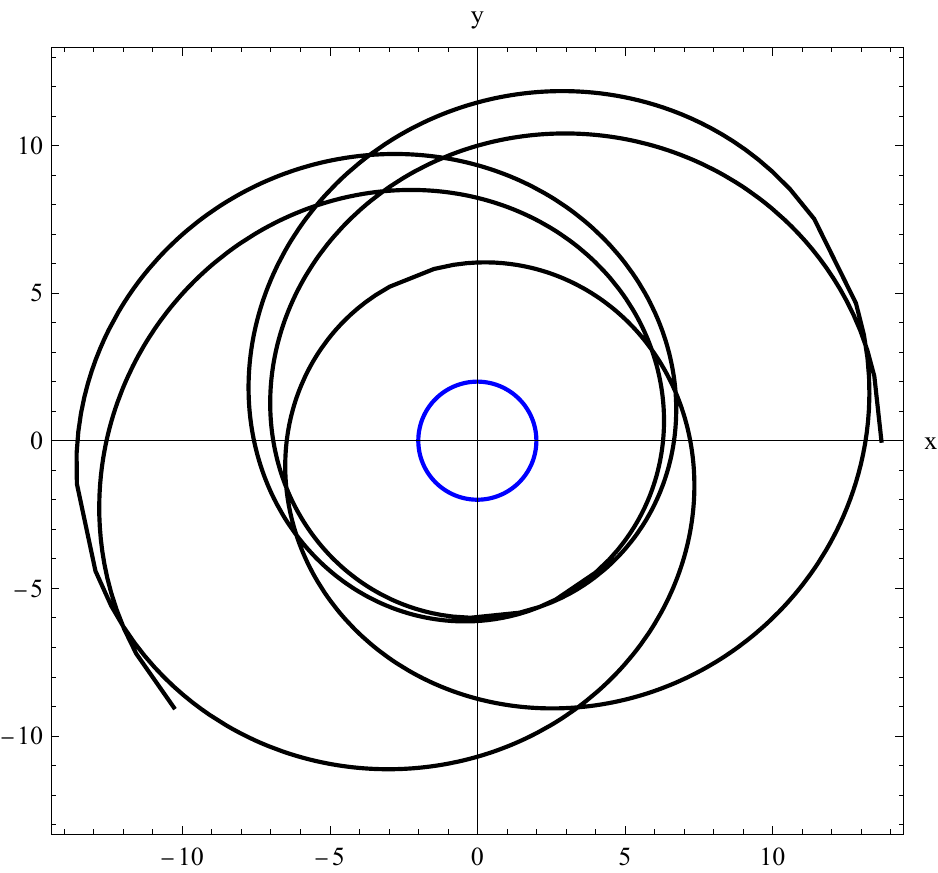}
  \caption{Parametric plot of the bounded trajectory in the equatorial plane of the particle with the energy $E=0.92
$ represented by the horizontal yellow line in Figure \ref{fig1}. The small blue circle represents the horizon.}
\label{fig3}
\end{figure}

One may notice in Figure \ref{fig1} that there is a region which allows periodic bounded trajectories, for the energy $V_{min} < E < 1$ represented by the horizontal yellow line. The parametric plot in the equatorial plane of such a closed trajectory in given in Figure \ref{fig3}.
For $E = V_{min}$, the particle is experiencing a stable circular motion.
The innermost stable circular orbits (ISCO) around the black hole for charged particles have been recently studied in \cite{Turimov:2020fme}, using a numerical approach.
A detailed discussion on the radial and angular motions with different types of trajectories can be found in \cite{Villanueva:2015kua}. For the Kepler-like orbits, the precession angles were also considered in \cite{Villanueva:2015kua}.

\begin{figure}
  \centering
  \includegraphics[width=0.45\textwidth]{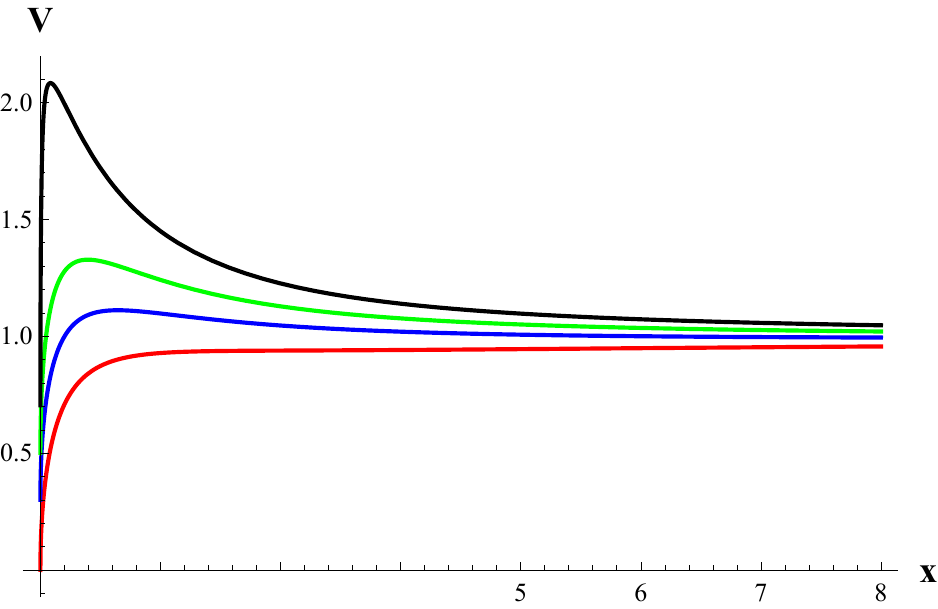} \hspace{0.2cm}
\includegraphics[width=0.45\textwidth]{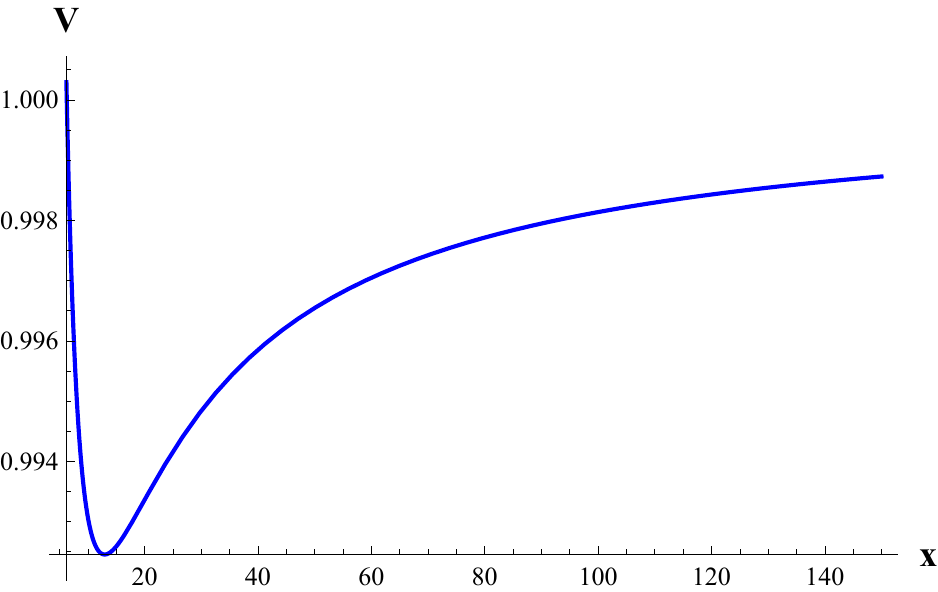} 
  \caption{{\it Left panel.} The potential (\ref{poteleff}) for $\tilde{Q}=0$ (the red plot), $\tilde{Q} =0.3$ (the blue plot), $\tilde{Q} =0.5$ (the green plot) and $\tilde{Q} =0.7$ (the black plot).
Other numerical values are: $M=1$, $q =1$, $\tilde{L}=1.7$.
{\it Right panel.} The effective potential (\ref{poteleff}), far from the horizon, for $\tilde{Q} =0.3$.} 
\label{fig4}
\end{figure}

The left panel in Figure \ref{fig4} is a graphical representation of the effective potential (\ref{poteleff}), for the angular momentum $\tilde{L}=1.7$.
Note that this value is less than $L/M = \sqrt{12}$ for which, in the Schwarzschild case (represented by the red plot), all the trajectories end up inside the black hole. The blue, green and black plots correspond to
$\tilde{Q} =0.3$, $\tilde{Q} =0.5$ and $\tilde{Q} =0.7$, respectively. One may notice, that once $\tilde{Q}$ is increasing, the potential gets a maximum value which is strongly increasing and comes closer to the horizon. Thus, the particle with $q>0$ and $E< V_{max}$ which is approaching this region coming from large distances will not fall into the black hole.

On the other hand, similarly to the Schwarzschild BH, the green and black plots do not show a minimum value of the potential. However, it is important to identify a finite domain where particles move on stable bound orbits, neither falling into the black hole nor escaping to infinity. In this respect, for $\tilde{Q} =0.3$ i.e. $r_0 = 0.36M$, a small potential well is formed, as it can be noticed in the right panel of Figure \ref{fig4}. Thus, $r_0 =0.36 M$ is a physically important value of the dilaton parameter for which the particles with $V_{min} < E<1$ can be trapped on bounded periodic orbits.
This value is much below the theoretical limit $r_0 =2M$ and agrees with the conclusions in \cite{Fernandez:2023kro}.

 Thus, one may conclude that the dilaton black hole charge $\tilde{Q}$ has a strong influence on the potential shape. There are values of the dilaton parameter for which the particles are moving on stable circular orbits or are trapped on bounded trajectories.
  
\subsection{The magnetic ansatz}

In the case of the magnetic black hole of charge $Q_m$ with the magnetic field generated by the vector potential component
\begin{equation}
A_{\varphi} = - Q_m \cos \theta \; ,
\label{Amag}
\end{equation}
the corresponding Lagrangean describing the motion of an electrically charged particle of unit mass and charge $q$ becomes:
\begin{equation}
{\cal L} =  \frac{1}{2} \left[ - f \dot{t}^2 + \frac{\dot{r}^2}{f} + p^2 ( \dot{\theta}^2+ \sin^2 \theta \dot{\varphi}^2 ) \right]  - q Q_m \cos \theta \dot{\varphi}.
\label{Lagmag}
\end{equation}
 Since the coordinates $t$ and $\varphi$ are cyclical the conserved energy becomes $E=f \dot{t}$ while the conservation of the total angular momentum along the axis $Oz$ corresponds to:
\begin{equation}
L_m =p^2\sin^2\theta\dot{\varphi}-qQ_m\cos\theta. 
\label{Lm}
\end{equation}

Generically, one expects that the motion of an electrically charged particles in the field of a magnetic monopole should be confined on the so-called Poincar\'e cones \cite{Poincare}. This is due to the $SO(3)$ symmetry of the system, even if the Lagrangian describing this motion is not manifestly exhibiting this $SO(3)$ symmetry, which is basically reflected in the geometry. As such, for the black hole geometry (\ref{metric}) there do exist three spacelike Killing vectors of the form:
\beqs
\xi_{(x)}^{\mu}&=&(0, 0, -\sin\varphi, -\cos\varphi \cot\theta), \nonumber\\
\xi_{(y)}^{\mu}&=&(0, 0, \cos\varphi, -\sin\varphi \cot\theta),\nonumber\\
\xi_{(z)}^{\mu}&=&(0, 0, 0, 1).
\eeqs
These Killing vectors can be subsequently used to define the components of the orbital angular momentum:
\beqs
L_x&=&\xi_{(x)}^{\mu}\dot{x}_{\mu}=-p^2\sin\varphi~\dot{\theta}-p^2\sin\theta\cos\theta\cos\varphi~\dot{\varphi},\nonumber\\
L_{y}&=&\xi_{(y)}^{\mu}\dot{x}_{\mu}=p^2\cos\varphi~\dot{\theta}-p^2\sin\theta\cos\theta\sin\varphi~\dot{\varphi},\nonumber\\
L_{z}&=&\xi_{(z)}^{\mu}\dot{x}_{\mu}=p^2\sin^2\theta~\dot{\varphi},
\label{orbitmom}
\eeqs
as well as a nontrivial Killing tensor of the second order:
\beqs
K^{\mu\nu}&=&\xi_{(x)}^{\mu}\xi_{(x)}^{\nu}+\xi_{(y)}^{\mu}\xi_{(y)}^{\nu}+\xi_{(z)}^{\mu}\xi_{(z)}^{\nu},
\eeqs
which can be used to construct the Carter constant:
\beqs
K&=&K_{\mu\nu}\dot{x}^{\mu}\dot{x}^{\nu}=p^4 \left[ \dot{\theta}^2 +  \sin^2 \theta \dot{\varphi}^2 \right].
\label{carter}
\eeqs
Note that the above expression of the Carter constant is related to the square of the magnitude of the orbital angular momentum: $K=L^2=L_x^2+L_y^2+L_z^2$.

If the motion were to be geodesic, these quantities will be all be conserved along the motion. However, in our case the motion is non-geodesic due to the presence of the electromagnetic forces acting on the electrically charged particle in the magnetic field of a magnetic monopole. Nonetheless, even if the components of the orbital angular momentum (\ref{orbitmom}) are not conserved during the motion, the Carter constant (\ref{carter}) is actually conserved, as we shall see bellow.

To this end one should note that the electrically charged particle has an angular momentum $\vec{S}$ with constant magnitude proportional to $-q Q_m$, which is directed along the radial direction, which connects it to the magnetic monopole in the origin, $\vec{S}=-qQ_m\hat{r}$. Defining now the total angular momentum $\vec{J}=\vec{S}+\vec{L}$ one can directly check that its components are all conserved along the motion:
\beqs
\frac{dJ_x}{d\tau}&=&\frac{dJ_y}{d\tau}=\frac{dJ_z}{d\tau}=0,
\eeqs
which means that the vector $\vec{J}$ is constant during the motion. To check the above equations one has to take into account the explicit equations of motion derived from (\ref{Lagmag}) for the $\theta$ and $\varphi$ variables. Note also that the constant $L_m$ in (\ref{Lm}) corresponds directly to the $z$-component of the total angular momentum $J_z=L_m$.

Using now the relation $\vec{J}  \cdot \hat{r} = -q Q_m$ one can see that the vector $\vec{r}$ is describing a cone whose axis is along $\vec{J}$, with the opening angle $\chi = 2 \alpha$, where
\begin{equation}
\cos \alpha = \frac{q Q_m}{J}
\label{conea}
\end{equation}
Also, the relation
\[
\vec{J} \cdot \vec{L} ={ L}^2 = J^2 - (q Q_m)^2 = const.
\]
is pointing out the existence of another cone, of constant angle $\beta$, with
\[
\cos \beta = \frac{L}{J} = \frac{L}{\sqrt{{L}^2 + ( q Q_m)^2}}
\]
on which the angular momentum $\vec{L}$ is moving around $\vec{J}$. Since $|\vec{J}|$ is constant this means that $L=|\vec{L}|$ and also the Carter constant in (\ref{carter}) is then a constant, as expected.

Since $\hat{r}$ is orthogonal on $\vec{L}$, it follows that $\alpha + \beta = \pi/2$. Thus, the particle's trajectory lies on a three-dimensional cone whose opening angle is completely determined by the total angular momentum and the charge combination $qQ_m$ as being $\chi = 2 \alpha$ with
\begin{equation}
\tan \alpha = \frac{{L}}{q Q_m}
\end{equation}
Finally, the angle between the direction of the total angular momentum $\vec{J}$ and the Oz axis is given by:
\beqs
\cos\delta&=&\frac{J_z}{J}=\frac{J_z}{K+(qQ_m)^2}.
\eeqs

Returning now to the characteristics of the motion along the angular directions, by using  
\begin{equation}
\dot{\varphi} = \frac{J_z + q Q_m \cos \theta}{p^2 \sin^2 \theta}
\label{dotphi}
\end{equation}
in the Carter constant (\ref{carter}) one obtains the important relation
\begin{equation}
\dot{\theta}^2 = \frac{1}{p^4} \left[ K -  \frac{ \left(  J_z + q Q_m \cos \theta  \right)^2}{\sin^2 \theta} \right]. 
\label{dottheta}
\end{equation}
Obviously, the right hand side should be a positive quantity and therefore one has to impose the relation
\begin{equation}
\left[ K + ( q Q_m)^2 \right] \cos^2 \theta + 2 q Q_m J_z \cos \theta + J_z^2 - K \leq 0,
\end{equation}
meaning $\cos \theta \in [ \cos \theta_1 , \, \cos \theta_2 ]$, where
\begin{equation}
\cos \theta_1 = \frac{- q Q_m J_z  - \sqrt{\Delta}}{K+( q Q_m)^2}
\; , \quad
\cos \theta_2 = \frac{- q Q_m J_z  + \sqrt{\Delta}}{K+(q Q_m)^2}
\label{theta12}
\end{equation}
The conditions $\cos \theta_{1,2} \in [-1 , 1]$ and 
\begin{equation}
\Delta = K \left[ K + (\varepsilon Q_m)^2 - J_z^2 \right]  \geq 0
\end{equation}
lead to the following the range of the angular momentum
\begin{equation}
- \sqrt{K + ( q Q_m)^2} < J_z <   \sqrt{K + ( q Q_m)^2}.
\label{rangeLz}
\end{equation}
Finally, using the relations (\ref{dotphi}) and (\ref{dottheta}) in the normalization condition  $g_{\mu\nu}\dot{x}^{\mu}\dot{x}^{\nu}=-1$, one obtains the equation describing the $r-$motion:
\begin{equation}
\dot{r}^2  = E^2 - f \left( 1 + \frac{K}{p^2} \right)  = E^2 - V.
\label{radmag}
\end{equation}
This has the same form as the one for uncharged particles with $K = J_z^2$, and the corresponding trajectories, in the equatorial plane, have been investigated by many authors, as for example in \cite{Blaga:2014spa}. The difference is that, in our case, the charged particle with the angular momentum in the range (\ref{rangeLz}) is following a trajectory which lies on a 3-dimensional Poincar\'e cone as we have seen above. (see also \cite{Dariescu:2023twk}, \cite{Lim:2021ejg} and \cite{Lim:2022qrt}).

On the other hand, since $K = L^2$ and $J^2 = L^2 + ( q Q_m)^2$, the relations in (\ref{theta12}) become
\[
\cos \theta_{1,2} = \frac{- q Q_m J_z  \pm L \sqrt{J^2 - J_z^2 }}{J^2}.
\]
These define the angles of two cones, which in general are not symmetric to the plane $xOy$ and the motion of the
test particle is confined in between them.

The periodic bounded trajectory for a particle whose orbital momentum is in the allowed range (\ref{radmag}) and the energy is in the classical region $V \leq E <1$, can be numerically obtained, using Maple or Mathematica. In this respect, we start with the Lagrangean (\ref{Lagmag}) and derive the system of Euler--Lagrange equations:
\begin{eqnarray}
&&
\ddot{r} = \frac{f^{\prime}}{2f} \dot{r}^2  - \frac{f^{\prime}E^2}{2f} +  f p p^{\prime} \dot{\theta}^2 + \frac{f p^{\prime} \left[ J_z + q Q_m \cos \theta \right]^2}{p^3 \sin^2 \theta}  \nonumber \\*
& & \ddot{\theta} = - \frac{2 p^{\prime}}{p} \dot{r} \dot{\theta} + \frac{\cos \theta \left[ J_z+ q Q_m \cos \theta \right]^2}{p^4 \sin^3 \theta} + \frac{q Q_m \left[ J_z + q Q_m \cos \theta \right]}{p^4 \sin \theta}
\label{lageq}
\end{eqnarray}
where $\dot{()}$ is the derivative with respect to $\tau$ and $()^{\prime}$ is the derivative with respect to $r$.

The equations in (\ref{lageq}) were solved numerically by using the Maple software and implementing a Runge-Kutta algorithm of 4th order. As a consistency check we used the first integral of motion given by (\ref{radmag}).

\begin{figure}
  \centering
  \includegraphics[width=0.5\textwidth]{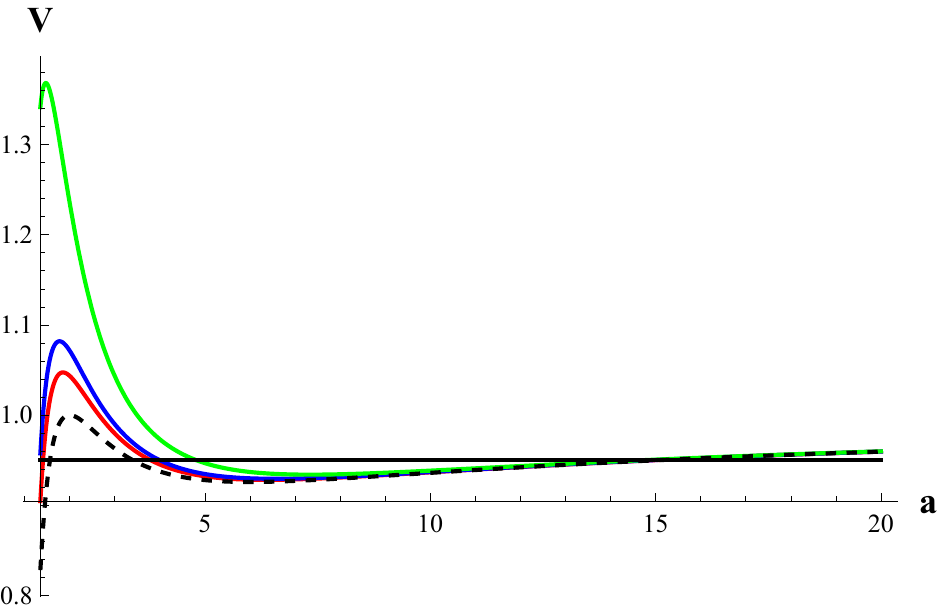}
  \caption{The potential defined in (\ref{radmag}) as a function of $x=r/(2M)$, for $a=0$ (the dashed black line), $a=0.08$ (the red curve), $a=0.125$ (the blue curve) and $a=0.32$ (the green plot). The black horizontal line represents the particle's energy $E= \sqrt{0.95}$. The other numerical values are: $2M=1$, $K=4$.}
  \label{fig5}
\end{figure}

\begin{figure}
  \centering
  \includegraphics[width=0.4\textwidth]{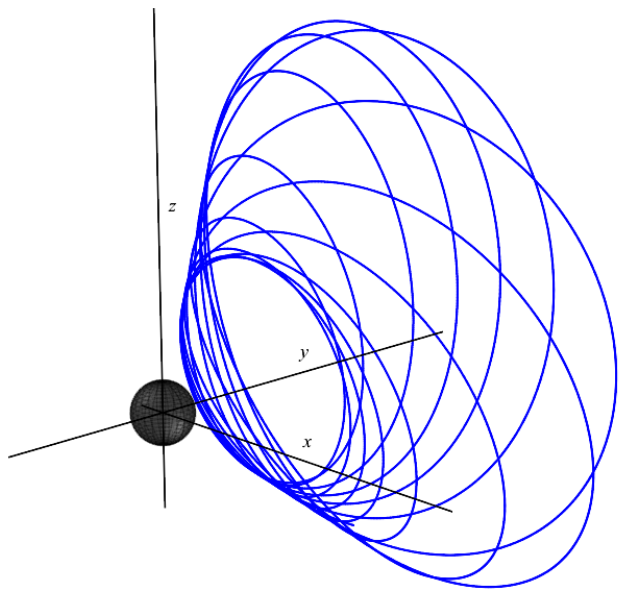} \hspace{0.2cm}
\includegraphics[width=0.4\textwidth]{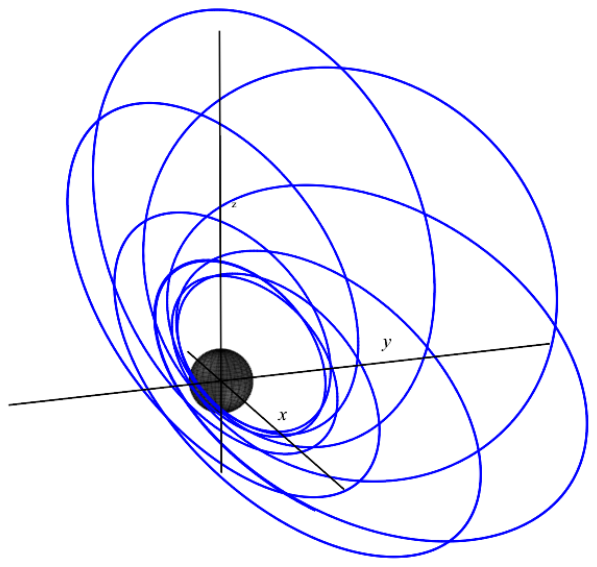}  \\
\caption{Orbits on cones for the potential represented in Figure \ref{fig5}. The solid sphere
represents the horizon. The left figure is for $Q_m = \sqrt{0.32}$ and the right one is for $Q_m = \sqrt{0.08}$. The other numerical values are: $2M=1$, $K=4$, $E=\sqrt{0.95}$, $q=3 \sqrt{2}$, $J_z=1.5$.} 
\label{fig6}
\end{figure}

In Figure \ref{fig5}, we represented the potential defined in (\ref{radmag}), as a function of the rescaled quantities $x=r/(2M)$ and $a = \tilde{Q}_m^2$. The black dashed plot corresponds to the Schwarzschild BH ($a=0$). As the charge $\tilde{Q}_m$ is increasing, the maximum of the effective potential is also increasing. The curves are approaching the  potential value $V=1$ for large values of the radial coordinate. A particle with the energy $E= \sqrt{0.95}$, represented by the black horizontal line will follow a periodic bounded orbit on a cone whose opening angle depends on the value of $\tilde{Q}_m$. Thus, when $\tilde{Q}_m$ is decreasing, the cone's angle becomes wider, as it can be seen in Figure \ref{fig6}. One may easily check that, when $\tilde{Q}_m \to 0$, one recovers the usual Schwarzschild case, with the periodic bounded orbit in the $xOy$ plane. 

\subsection{Circular motion on the Poincar\'e cone}

At the end of this section, let us briefly mention the trajectory of constant $r$, which can be named as being a circular orbit.
The term ‘circular motion’ used here was explained in \cite{Lim:2021ejg}.
Thus, the constant-$r$ trajectories
may lie on a sphere and, at the same time, on the Poincar\'e cone and the intersection between these two is, again, a circle.
According to the relation (\ref{conea}), by changing the angular momentum, one is able to change both the opening angle and the orientation of the cone on which the circular orbit is situated.

As it is known, a circular orbit of radius $r_c$ must satisfy the usual conditions:
\[
V (r _c) = E^2 \; ; \quad V^{\prime}  (r_c) = 0
\]
where $V(r)$ is defined in (\ref{radmag}) by:
\[
V(r) =  1 - \frac{2M}{r} + \frac{K}{p^2} - \frac{2 MK}{r p^2}.
\]
Thus, the energy of the corresponding particle must have the expression:
\[
E^2  = \frac{(2M - r_c)^2 (2r_c-r_0)}{r_c [2r_c^2 - (6M+r_0)r_c+4Mr_0]} 
\]
while the Carter's constant $K = L^2$ is:
\[
K = \frac{2M  r_c (r_c-r_0)^2}{2r_c^2 - (6M+r_0)r_c+4Mr_0}.
\]
If we impose that the above quantities are positive, \textit{i. e.} $2r_c^2 - (6M+r_0)r_c+4Mr_0 >0$,
we obtain the allowed range of the circular orbit radius
\beqs
r_c > r_* = \frac{6M+r_0 + \sqrt{36M^2-20M r_0 + r_0^2}}{4}.
\label{rmin}
\eeqs
The minimum value $r_*$ is depending on the dilaton's parameter $r_0$ and will take values in the range $r_* \in ( 2M , 3M)$. Thus, for $r_0 =0$, we recover the unstable circular orbit of the Schwarzschild black hole, $r_c =3M$. While $r_0$ is increasing, the value $r_*$ is decreasing to $r_* =2M$ for $r_0 =2M$.

\begin{figure}
  \centering
  \includegraphics[width=0.45\textwidth]{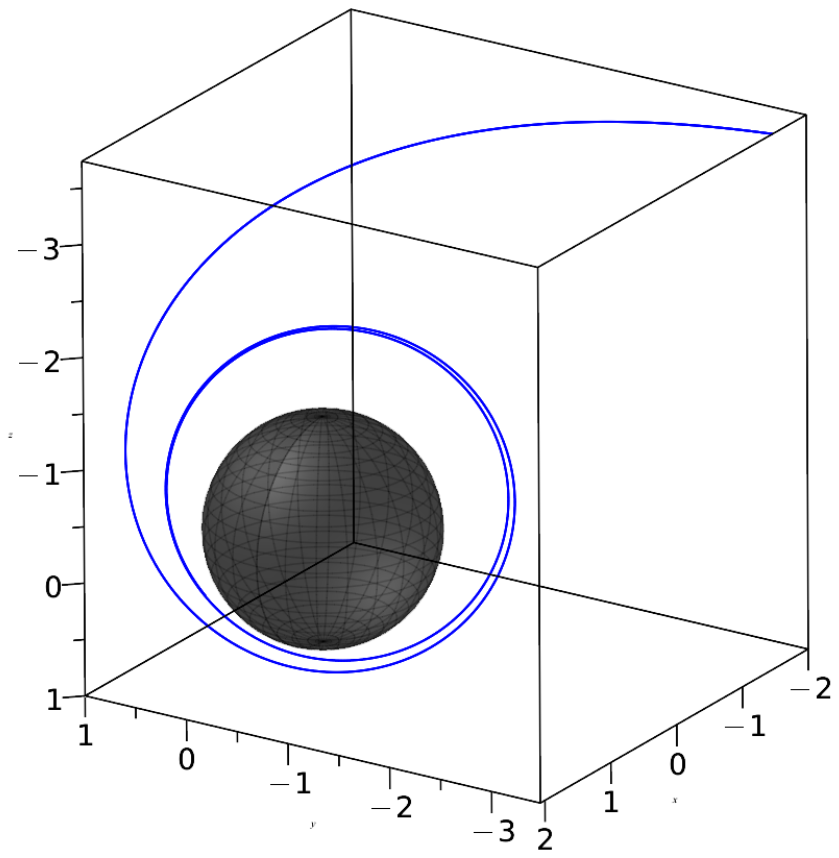} \hspace{0.2cm}
\includegraphics[width=0.45\textwidth]{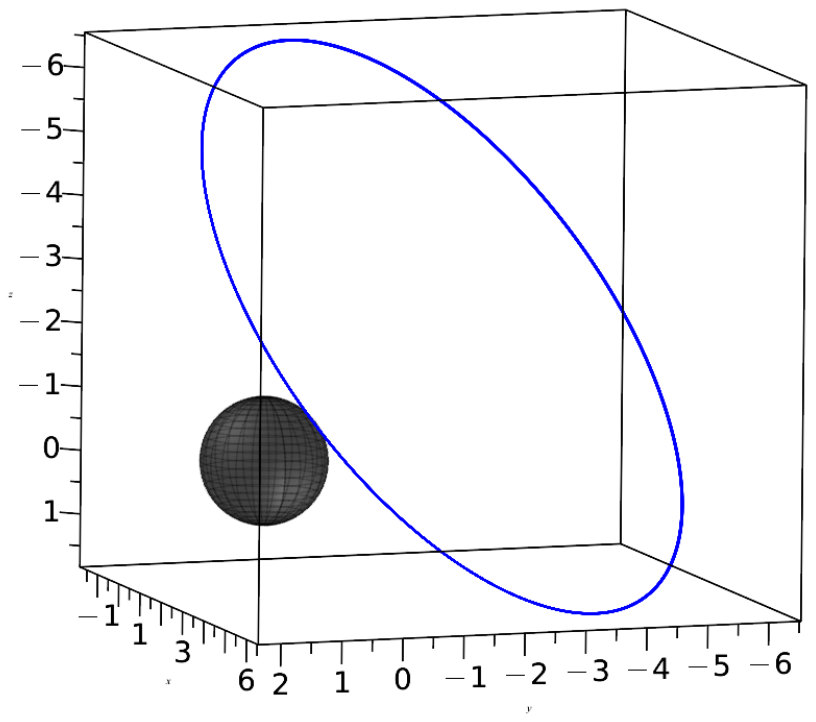}  
\caption{{\it Left panel.} Unstable circular orbit with $r_{c1} = 1.7747$ and $E_1^2 = 1.08182$. {\it Right panel.} Stable circular orbit with $r_{c2} = 6.553$ and $E_2^2 = 0.9294$. The solid sphere
represents the horizon. The other numerical values are: $2M=1$, $Q_m = \sqrt{0.125}$, $q  =3 \sqrt{2}$, $J_z=1.5$.} 
\label{fig7}
\end{figure}

For the potential with $a= Q^2_m = 0.125$, represented in Figure \ref{fig5} by the blue line, one has an unstable circular orbit and a stable one corresponding to the maximum and minimum values of the potential. These trajectories are represented in Figure \ref{fig7}.

 One may conclude by saying that the radius of the circular orbit and its corresponding range are depending on the dilaton parameter. Once $Q_m$ is decreasing, the particle is moving on a circular orbit which lies on a cone whose angle is wider until it reaches the $xOy$ plane for $Q_m \to 0$.

\section{The $U(1)$ gauge-covariant Klein--Gordon equation}

\subsection{The analytical solution in the electric case}

In the case of a charged particle characterized by $q >0$ and $m_0$, the
Klein--Gordon equation has the general expression:
\begin{equation}
g^{ij} D_i D_j \Phi - m_0^2 \Phi \, = \, 0, 
\label{KGeq}
\end{equation}
where the gauge derivatives defined as:
\[
D_i = \partial_i  - i q A_i ,
\]
contain the electromagnetic potential component $A_t$ given in (\ref{At}).
Thus, the extended form of the equation (\ref{KGeq}) is
\begin{eqnarray}
& &
\frac{1}{p^2} \left \lbrace \frac{\partial \;}{\partial r} \left[ p^2 f \frac{\partial \Phi}{\partial r} \right] + \frac{1}{\sin \theta} \frac{\partial \;}{\partial \theta} \left( \sin \theta \frac{\partial \Phi}{\partial \theta}   \right) + \frac{1}{\sin^2 \theta} \frac{\partial^2 \Phi}{\partial \varphi^2} \right \rbrace
+ \frac{1}{f} \left[ i \frac{\partial \;}{\partial t} - \frac{qQ}{r} \right]^2 \Phi 
- m_0^2 \Phi = 0. \nonumber
\end{eqnarray}
With the variables separation:
\begin{equation}
\Phi = R (r) Y_l^m ( \theta , \varphi ) \, e^{-i \omega t} \, ,
\end{equation}
where $Y_l^m$ are the usual spherical functions, one obtains the following radial equation:
\begin{eqnarray}
\frac{d \;}{dr} \left[ p^2 f \frac{dR}{dr} \right] + \left[ \frac{p^2}{f} \left( \omega - \frac{qQ}{r} \right)^2 - m_0^2 p^2 - l(l+1)  \right]
R = 0,
\label{radialeq}
\end{eqnarray}
whose solutions can be obtained using Maple. Up to some integration constants, these are given by the Heun confluent functions \cite{Heun}, as: 
\begin{eqnarray}
R(r) & = & C_{1,2} e^{ikr} \left( r - 2M \right)^{\beta /2} HeunC \left[ \alpha, \, \beta , \, \gamma , \, \delta , \, \eta , \; z \right] ,
\label{radR}
\end{eqnarray}
with $k = \sqrt{\omega^2 - m_0^2}$, the variable is:
\[
z = \frac{2M - r}{2M - r_0} = \frac{1-x}{1-2a}
\]
and the Heun function's parameters are:
\begin{eqnarray}
& &
\alpha = -2ik (1 -2a ) \; , \quad \beta = \pm 2i ( \omega - q \tilde{Q} ) \; , \quad \gamma = 0 \; , \nonumber \\*
& &
\delta = - (1-2a) \left( 2 \omega^2 - m_0^2 - 2 \omega q \tilde{Q} \right) \; , \quad \eta = - \delta -l(l+1),
\label{Hparam}
\end{eqnarray}
where $a = r_0/(4M) = \tilde{Q}^2$ is in the physical range $a \in [0, 0.5]$. In the followings, we shall consider the value $\beta = -2i (\omega - q \tilde{Q} )$. The solution with $\beta = + 2i (\omega - q \tilde{Q} )$ corresponds to $\omega \to - \omega$ and $q \to -q$.

\begin{figure}
\centering
  \includegraphics[width=0.7\textwidth]{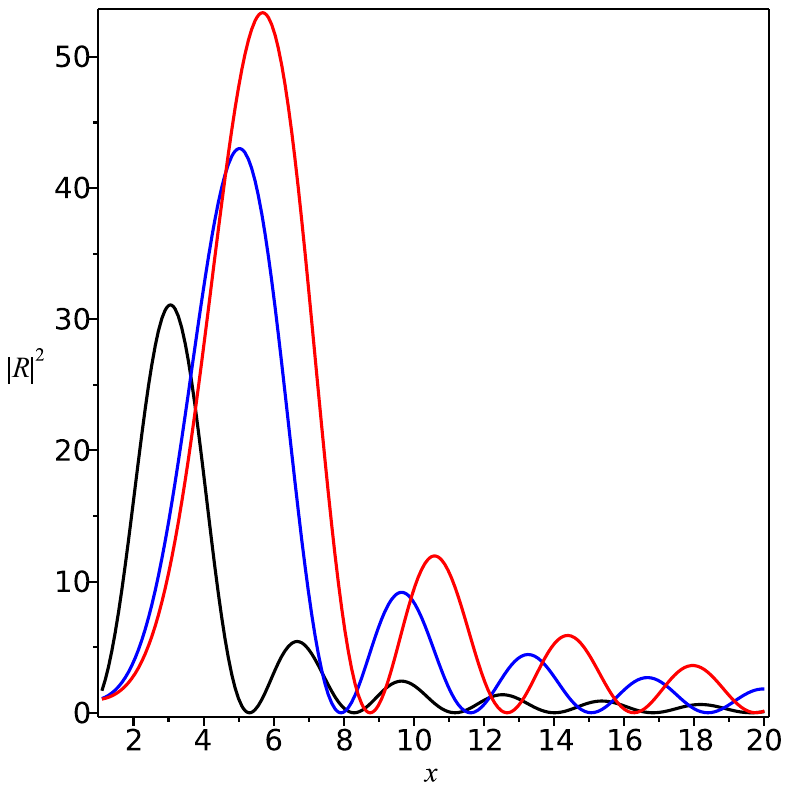}
  \caption{The absolute value of the radial function (29), as a function of $x = r/(2M)$, for different values of $a = \tilde{Q}^2$. Here $M=1$, $\omega=0.5$, $m_0=0.3$, $q=1$ and $l=1$.}
  \label{fig8}
\end{figure}

The absolute value of the radial function (\ref{radR}) is represented in Figure \ref{fig8}, as a function of the rescaled coordinate $x = r/(2M)$, for different values of the dilaton parameter $a = \tilde{Q}^2$. 
The density probability is given by the square modulus of the Heun confluent function. One may notice that $\left| R \right|^2 =1$ on the horizon, it gets a series of local decreasing maxima and is vanishing for large $r$'s. The black hole's charge is affecting the Heun function parameters and its variables. Thus, the maxima have higher values as $\tilde{Q}$ increases and they are shifted to larger radial distances.

In order to study the scattering of the charged scalar field in the background of this dilatonic
black hole, one has to put the radial equation in a Schrodinger-like form. Thus,
with the change of function $R(r) = F(r)/p$,
and using the tortoise coordinate:
\[
\frac{dr_*}{dr} = \frac{1}{f(r)},
\]
the radial equation (\ref{radR}) gets the Schrodinger-like expression:
\[
\frac{d^2 F}{dr_*^2} + \left \lbrace \left( \omega - \frac{qQ}{r} \right)^2  - f \left[ \frac{l(l+1)}{p^2} + m_0^2 \right] - \frac{f}{p} \frac{d \;}{dr} \left[ f p^{\prime} \right]
\right \rbrace F = 0
\]
In the eikonal approximation, i.e. large values of $l$, and for $m_0 =1$, this turns into the simple expression:
\begin{equation}
\frac{d^2 F}{dr_*^2} + \left[ \left( \omega - \frac{qQ}{r} \right)^2 - f \left( \frac{l^2}{p^2} + 1 \right)  \right] F = 0,
\label{schwar}
\end{equation}
where one can notice the same form of the potential as the one in (\ref{radialel}). The potential depends on the particle's charge and angular momentum as well as on the black hole's parameters $M$ and $Q$.

\subsection{Energy spectrum}

The confluent Heun function has a polynomial form for the following relation between its parameters \cite{Heun} - \cite{Birkandan:2017rdp}:
\begin{equation}
\frac{\delta}{\alpha} = - \left[ n+1 + \frac{\beta+\gamma}{2} \right],
\label{poly}
\end{equation}
and the general solution of (\ref{KGeq}) can be written as:
\begin{equation}
\Phi_{nlm} = \Sigma_{nlm} R_{nl} (r) Y_l^m ( \theta , \varphi ) e^{-i \omega_n t},
\label{Phigen}
\end{equation}
where $R_{nl}$ is given in (\ref{radR}).
For the parameters (\ref{Hparam}), the above condition leads to the following cubic equation:
\begin{eqnarray}
& &
-8i (n+1) \omega^3 + 4 (n+1) \left[ n+1+2iq \tilde{Q} \right] \omega^2 + 4 i m_0^2  \left[ 2(n+1)+iq \tilde{Q} \right] \omega \nonumber \\*
& & + m_0^2\left[ m_0^2- 4 (n+1+i q  \tilde{Q} )^2 \right] = 0,
\label{cubiceq}
\end{eqnarray}
where the complex energy is $\omega = \omega_I + i \omega_R$. 

\begin{figure}
  \centering
  \includegraphics[width=0.7\textwidth]{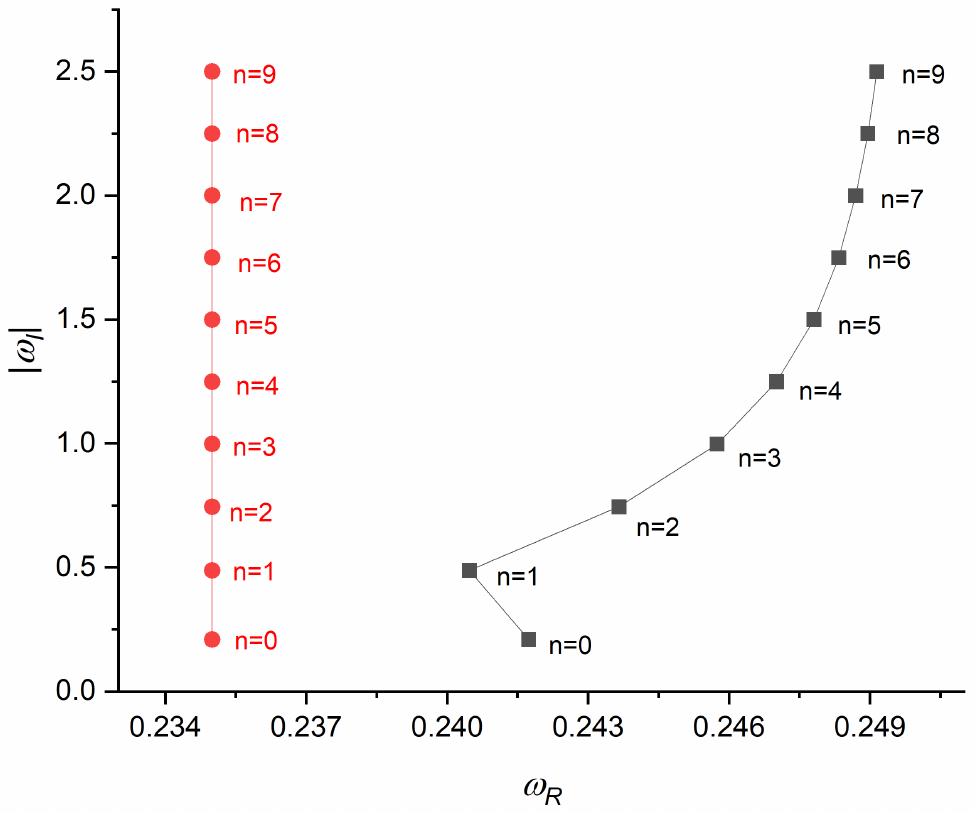}
  \caption{The real and imaginary parts of $\omega$, solutions of (34), for different $n-$values. The numerical values are: $\tilde{Q} = 0.25$, $m_0=0.3$, $q=1$.}
  \label{fig9}
\end{figure}

Among the three roots of (\ref{cubiceq}), we choose, for each value of $n$, the complex one with negative imaginary part so that the function (\ref{Phigen}) decreases exponentially for $t \to \infty$. The real part $\omega_R$ and the absolute value of the imaginary part $\omega_I$ are represented in Figure \ref{fig9}, for different $n$ values. The imaginary part takes equally spaced numerical values which are increasing with $n$. These are projected on the red vertical line.

In the massless case, the radial function corresponding to the uncharged boson has the same expression as in (\ref{radR}), with $k = \omega$, while the Heun function's parameters have the simple expressions
\begin{equation}
\alpha_0 = 2 i \omega (1 - 2a) \; , \quad \beta_0 = \pm \, 2i \omega  \; , \quad \gamma = 0 \; , \quad \delta_0 = - 2 \omega^2 (1- 2a) .
\label{param2}
\end{equation}

If one imposes the relation (\ref{poly}), the Heun confluent functions get a polynomial form.
By using the expressions of the parameters (\ref{param2}), in the condition (3\ref{poly}), we get the same energy
quantization law as in the Schwarzschild case \cite{Vieira:2016ubt}:
\begin{equation}
\omega =  - i \frac{( n + 1)}{4M} \, .
\label{schw}
\end{equation}

\subsection{The magnetic ansatz}

For the magnetic black hole, we use the electromagnetic potential component (\ref{Amag}) in the expression of the gauge derivative, $D_{\varphi} = \partial_{\varphi} + iq Q_m \cos \theta$, and the Klein--Gordon equation (\ref{KGeq}) has the explicit form:
\begin{eqnarray}
& &
\frac{1}{p^2} \left \lbrace \frac{\partial \;}{\partial r} \left[ p^2 f \frac{\partial \Phi}{\partial r} \right] + \frac{1}{\sin \theta} \frac{\partial \;}{\partial \theta} \left( \sin \theta \frac{\partial \Phi}{\partial \theta}   \right) + \frac{1}{\sin^2 \theta} \left[ \frac{\partial \;}{\partial \varphi} +i q Q_m \cos \theta \right]^2 \Phi  \right \rbrace
\nonumber \\*
& & - \frac{1}{f} \frac{\partial^2 \Phi}{\partial t^2}  
- m_0^2 \Phi = 0 \; .
\end{eqnarray}
With the variables separation:
\begin{equation}
\Phi = G (r) T( \theta ) e^{im \varphi} \, e^{-i \omega t} \, ,
\end{equation}
one obtains the following system of decoupled equations:
\begin{eqnarray}
& & \frac{d \;}{dr} \left[ p^2 f \frac{dG}{dr} \right] + \left[ \frac{p^2 \omega^2}{f} - m_0^2 p^2 - \lambda  \right]
G = 0, \nonumber \\*
& & \frac{1}{\sin \theta} \frac{d \;}{d \theta} \left[ \sin \theta \frac{dT}{d \theta} \right]  + \left[ \lambda - \frac{(m + qQ_m \cos \theta)^2}{\sin^2 \theta} \right] T =0
\; .
\end{eqnarray}

The azimuthal equation is satisfied by hypergeometric functions, while the solutions to the radial equation
can be written in terms of Heun confluent functions as 
\begin{eqnarray}
G(r) & = & C_{1,2} e^{ikr} \left( r - 2M \right)^{\beta /2} HeunC \left[ \alpha, \, \beta , \, \gamma , \, \delta , \, \eta , \; z \right] ,
\end{eqnarray}
with the same variable as in (\ref{radR}) and the parameters (\ref{Hparam}) written for $ q \tilde{Q}=0$, namely:
\begin{eqnarray}
& &
\alpha = -2ik (1 - 2a ) \; , \quad \beta = \pm 2i  \omega \; , \quad \gamma = 0 \; , \nonumber \\*
& &
\delta = - (1-2a )(2 \omega^2 -  m_0^2 ) \; , \quad \eta = - \delta -l(l+1).
\end{eqnarray}
The relation (\ref{cubiceq}) becomes:
\[
-8i (n+1) \omega^3 + 4 (n+1)^2 \omega^2 + 8 i (n+1)m_0^2 \omega 
+ m_0^2\left[ m_0^2- 4 (n+1 )^2 \right] = 0,
 \]
which is the same as the one corresponding to a massive particles in the Schwarzschild black hole \cite{Vieira:2016ubt}. 

\section{Conclusions}

In the present work we considered the background geometry of the GMGHS black hole. This geometry is a solution of the low-energy field description of the heterotic string theory compactified down to four dimensions on a six-dimensional torus. It can be described using the black hole mass $M$, the black hole charge $Q$ (in the electric or the magnetic case) and the asymptotic value of the dilaton field $\phi_0$. For simplicity, we have considered here a vanishing asymptotic value of the dilaton field, $\phi_0=0$.

Note that there also exists a `secondary hair', as described by the dilaton charge $D$, which is in this case proportional to the dilaton parameter $r_0=\frac{Q^2}{M}$.

Since this black hole is charged, the motion of electrically charged particles has new interesting features (as opposed to the uncharged test particles). This is due to the extra interaction between the particle and the electromagnetic field in the background.

In our work we made use of a Lagrangean approach to derive the corresponding equations of motion in both the electrically charged case, as well as for the magnetically charged black hole. In the electric case our results can be compared to those derived by Villanueva and Olivares who solved the equations of motion using the Hamilton--Jacobi method \cite{Villanueva:2015kua}. For charged particles moving in an electric field, the effective potential (\ref{potel}) is represented in Figure \ref{fig1}. The shape of the potential depends on the model's parameters and it leads to different types of trajectories, as it can be noticed in the Figures \ref{fig2} and \ref{fig3}.
Using the potential represented in Figure \ref{fig4}, we have shown that bounded periodic trajectories can be obtained for the preferred value $\tilde{Q} =0.3$, i.e. $r_0 =0.36$.
Such a conclusion agrees with the experimental bounds on $r_0$ imposed by astrophysical observations. For example, the optical continuum spectrum of the quasars has provided $r_0 =0.2$ as a preferred value \cite{Banerjee:2020qmi}. Recently, the measured shadow diameter for M87* and Sgr A* proposed as the allowed interval $r_0 \in [ 0.1 , 0.4]$ \cite{Sahoo:2023czj}.\footnote{We are using the unit mass $M=1$.}

The case of a magnetically charged black hole was addressed in Section $2.3$, where we described the main characteristics of the motion of an electrically charged particle in this background. We showed directly the conservation of the total angular momentum $\vec{J}$, which will define the direction of the axis of the Poincar\'e cone on which the motion of the particle takes place. In particular, the angular momentum component $J_z$ of the charged particle must be in the range (\ref{rangeLz}). Once $Q_m$ is increasing, the range (\ref{rangeLz}) is also increasing. The values $\theta_1$ and $\theta_2$ in the relation (\ref{theta12}), which correspond to $\dot{\theta}=0$, define 
the angles of two cones that bound the motion of
test particles. One may notice that for nonzero $J_z$ and $Q_m$ the two cones are not symmetric with respect to the equatorial plane. 

The particle's trajectory is not only constrained by these two cones but it also lies on a three-dimensional Poincar\'e cone. As it can be seen in equation (\ref{conea}), the opening angle of this cone is depending on the angular momentum and on the magnetic charge $Q_m$. As the black hole's charge is decreasing, the cone's angle becomes wider and the cone becomes much flatter as it can be noticed in Figure \ref{fig6}.

In particular, our results in the magnetic case can be compared to those obtained for instance in \cite{Gonzalez:2017kxt}. While the full equations of motion for charged test particles in the magnetic case can be integrated exactly in terms of Weierstrass' s special functions, in our Lagrangean approach one can easily show that the motion of electrically charged particles in presence of a magnetically charged black hole is confined to Poincar\'e cones of various angles. In particular, using our method it is easy to determine the characteristics of the Poincar\'e cones on which the motion of electrically charged particles is bounded to and their dependence on the physical quantities describing the GMGHS geometry and the charged test particle. In particular, we also investigate the existence of circular orbits located at the intersection of a Poincar\'e cone with a sphere.

The second part of our paper is dedicated to finding exact solutions of the charged Klein-Gordon equation in the GMGHS background.
The role of the scalar field has been extensively investigated in different contexts and analytical solutions to the Gordon equation are particularly important \cite{Dariescu:2010zz}. In the Einstein-Maxwell-dilaton theory, the scattering and absorption of a scalar field impinging on a charged black hole 
have been discussed in \cite{Richarte:2021fbi}. Recently, Heun-type solutions of the wave equation of scalar particles for different types of black holes have been derived in \cite{Birkandan:2017ypz}, \cite{Birkandan:2020ivm}.

In our work, we solved the Klein--Gordon equation in the GMGHS background described by the line element (\ref{metric}) and expressed its analytical solution in terms of Heun confluent functions. As it can be noticed in Figure \ref{fig8}, the value of the black hole's charge is strongly affecting the behavior of the radial amplitude function.
Thus, when $\tilde{Q}$ is increasing, the maximum values become more prominent and they move to larger $r-$values.

Even though the radial equation (\ref{radialeq}) and its solutions are the same as the ones derived in \cite{Vieira:2018djw}, the energy spectrum and its dependence on the model's parameters are quite different, the conclusions in \cite{Vieira:2018djw} being based on a quartic equation. In our case, we have derived the cubic equation (\ref{cubiceq}) and, among the three solutions, we have chosen, for each value of $n$, the complex root with $\omega_I <0$, so that the function (\ref{Phigen}) decreases exponentially for $t \to \infty$. 

The presence of the dilaton has a strong influence on the quasi-spectrum and we are interested in the imaginary part of $\omega$, which is characterizing the decay rate of the oscillations. As it can be noticed in Figure \ref{fig9}, $\omega_I$ takes almost equally spaced numerical values which are increasing with increasing $n$. These values are projected on the vertical thin red line. The sign of the test particle's charge, $q$, is affecting only the real part of $\omega$ which changes the sign, while the imaginary part remains the same. 

For massless bosons, the situation is similar to the Schwarzschild case \cite{Vieira:2016ubt}, but the parameters of the Heun confluent function and the variable are affected by the black hole's charge. However, the equally spaced energy spectrum (\ref{schw}) remains the same as for the Schwarzschild black hole. For fermions, a similar analysis has been performed in \cite{Dariescu:2018rrr}. 

For a direct transition from the Klein-Gordon equation to the classical limit of particles moving along
timelime geodesics, we have shown that, in the eikonal limit, the effective potential in (\ref{radialel}) has the same form as the one given in the Schrodinger-like equation (\ref{schwar}).

As avenues for further work, it might be interesting to extend the above analysis to the more general case of the charged dilatonic black hole with an arbitrary dilatonic coupling constant. In particular, for a particular coupling constant one can recover the black hole geometry obtained by Kaluza-Klein compactification from the corresponding five-dimensional black hole. It might be of interest how the black hole properties in five dimensions are related to those of the four-dimensional Kaluza-Klein compactification \cite{Stelea:2011fj}. Work on this issues is in progress and it will be reported elsewhere.

{\bf Conflicts of Interest.}
The authors declare that they have no conflicts of interest.

{\bf Data Availability.}
No data were used to support this study.


\begin{thebibliography}{99}

\bibitem{Mizuno:2018lxz}
Y.~Mizuno, Z.~Younsi, C.~M.~Fromm, O.~Porth, M.~De Laurentis, H.~Olivares, H.~Falcke, M.~Kramer and L.~Rezzolla,
Nature Astron. \textbf{2}, no.7, 585-590 (2018)
doi:10.1038/s41550-018-0449-5
[arXiv:1804.05812 [astro-ph.GA]].

\bibitem{Garfinkle:1990qj}
D.~Garfinkle, G.~T.~Horowitz and A.~Strominger,
Phys. Rev. D \textbf{43}, 3140 (1991)
[erratum: Phys. Rev. D \textbf{45}, 3888 (1992)]
doi:10.1103/PhysRevD.43.3140

\bibitem{GM} 
Gibbons, G.~W. \& Maeda, K.-I.\ 1988, Nuclear Physics B, 298, 741. doi:10.1016/0550-3213(88)90006-5

\bibitem{Sen:1992ua}
A.~Sen,
Phys. Rev. Lett. \textbf{69}, 1006-1009 (1992)
doi:10.1103/PhysRevLett.69.1006
[arXiv:hep-th/9204046 [hep-th]].

\bibitem{Narang:2020bgo}
A.~Narang, S.~Mohanty and A.~Kumar,
[arXiv:2002.12786 [gr-qc]].

\bibitem{GRAVITY:2020gka}
R.~Abuter \textit{et al.} [GRAVITY],
Astron. Astrophys. \textbf{636}, L5 (2020)
doi:10.1051/0004-6361/202037813
[arXiv:2004.07187 [astro-ph.GA]].

\bibitem{Fernandez:2023kro}
R.~F.~Fern\'andez, R.~Della Monica and I.~de Martino,
JCAP \textbf{08}, 039 (2023)
doi:10.1088/1475-7516/2023/08/039
[arXiv:2306.06937 [gr-qc]].

\bibitem{Banerjee:2020qmi}
I.~Banerjee, B.~Mandal and S.~SenGupta,
Mon. Not. Roy. Astron. Soc. \textbf{500}, no.1, 481-492 (2020)
doi:10.1093/mnras/staa3232
[arXiv:2007.13980 [gr-qc]].

\bibitem{Sahoo:2023czj}
S.~K.~Sahoo, N.~Yadav and I.~Banerjee,
Phys. Rev. D \textbf{109}, no.4, 044008 (2024)
doi:10.1103/PhysRevD.109.044008
[arXiv:2305.14870 [gr-qc]].

\bibitem{Tripathi:2021rwb}
A.~Tripathi, B.~Zhou, A.~B.~Abdikamalov, D.~Ayzenberg and C.~Bambi,
JCAP \textbf{07}, 002 (2021)
doi:10.1088/1475-7516/2021/07/002
[arXiv:2103.07593 [astro-ph.HE]].

\bibitem{Banerjee:2020ubc}
I.~Banerjee, B.~Mandal and S.~SenGupta,
Phys. Rev. D \textbf{103}, no.4, 044046 (2021)
doi:10.1103/PhysRevD.103.044046
[arXiv:2007.03947 [gr-qc]].

\bibitem{Feng:2023iha}
H.~Feng, Y.~Wu, R.~J.~Yang and L.~Modesto,
Phys. Rev. D \textbf{109}, no.6, 063014 (2024)
doi:10.1103/PhysRevD.109.063014
[arXiv:2301.02779 [astro-ph.HE]].

\bibitem{Feng:2024iqj}
H.~Feng and R.~J.~Yang,
[arXiv:2403.18541 [gr-qc]].

\bibitem{An:2017hby}
J.~An, J.~Peng, Y.~Liu and X.~H.~Feng,
Phys. Rev. D \textbf{97}, no.2, 024003 (2018)
doi:10.1103/PhysRevD.97.024003
[arXiv:1710.08630 [gr-qc]].


\bibitem{Wu:2001xh}
S.~Q.~Wu and X.~Cai,
[arXiv:gr-qc/0107037 [gr-qc]].

\bibitem{Vieira:2018hij}
H.~S.~Vieira and V.~B.~Bezerra,
Chin. Phys. C \textbf{43}, no.3, 035102 (2019)
doi:10.1088/1674-1137/43/3/035102
[arXiv:1811.06129 [gr-qc]].


\bibitem{Wu:2021pgf}
X.~Wu and X.~Zhang,
Universe \textbf{8}, no.11, 604 (2022)
doi:10.3390/universe8110604
[arXiv:2112.11066 [gr-qc]].

\bibitem{Cardoso:2008bp}
V.~Cardoso, A.~S.~Miranda, E.~Berti, H.~Witek and V.~T.~Zanchin,
Phys. Rev. D \textbf{79}, no.6, 064016 (2009)
doi:10.1103/PhysRevD.79.064016
[arXiv:0812.1806 [hep-th]].

\bibitem{Blaga:2014spa}
C.~Blaga,
Serb. Astron. J. \textbf{190}, 41 (2015)
doi:10.2298/SAJ1590041B
[arXiv:1407.1504 [gr-qc]].


\bibitem{Pradhan:2012id}
P.~P.~Pradhan,
Int. J. Mod. Phys. D \textbf{24}, 1550086 (2015)
doi:10.1142/S0218271815500868
[arXiv:1210.0221 [gr-qc]].


\bibitem{Fernando:2011ki}
S.~Fernando,
Phys. Rev. D \textbf{85}, 024033 (2012)
doi:10.1103/PhysRevD.85.024033
[arXiv:1109.0254 [hep-th]].

\bibitem{Soroushfar:2016yea}
S.~Soroushfar, R.~Saffari and E.~Sahami,
Phys. Rev. D \textbf{94}, no.2, 024010 (2016)
doi:10.1103/PhysRevD.94.024010
[arXiv:1601.03143 [gr-qc]].

\bibitem{Villanueva:2015kua}
J.~R.~Villanueva and M.~Olivares,
Eur. Phys. J. C \textbf{75}, no.11, 562 (2015)
doi:10.1140/epjc/s10052-015-3794-x
[arXiv:1510.08340 [gr-qc]].

\bibitem{Gonzalez:2017kxt}
P.~A.~Gonz\'alez, M.~Olivares, E.~Papantonopoulos, J.~Saavedra and Y.~V\'asquez,
Phys. Rev. D \textbf{95}, no.10, 104052 (2017)
doi:10.1103/PhysRevD.95.104052
[arXiv:1703.04840 [gr-qc]].

\bibitem{Poincare}
H.~Poincar\'e,
Compt. Rendus 123 (1896) 530–533.

\bibitem{Mondal:2020pop}
M.~Mondal, A.~K.~Yadav, P.~Pradhan, S.~Islam and F.~Rahaman,
Int. J. Mod. Phys. D \textbf{30}, no.12, 2150095 (2021)
doi:10.1142/S0218271821500954
[arXiv:2009.03265 [gr-qc]].

\bibitem{Heun} 
A. Ronveaux,
{\it Heun's differential equations},
 Oxford Science Publications
(The Clarendon Press Oxford University Press,1995).
%
\\
%
S. Yu. Slavianov and W. Lay, 
{\it Special Functions: A Unified Theory Based on Singularities},	
Oxford University Press (2000).

\bibitem{Birkandan:2017rdp}
T.~Birkandan and M.~Horta\c{c}su,
EPL \textbf{119}, no.2, 20002 (2017)
doi:10.1209/0295-5075/119/20002
[arXiv:1704.00294 [math-ph]].

\bibitem{Birkandan:2022zny}
T.~Birkandan,
Gen. Rel. Grav. \textbf{55}, no.8, 88 (2023)
doi:10.1007/s10714-023-03134-3
[arXiv:2204.10627 [gr-qc]].

\bibitem{Vieira:2016ubt}
H.~S.~Vieira and V.~B.~Bezerra,
Annals Phys. \textbf{373}, 28-42 (2016)
doi:10.1016/j.aop.2016.06.016
[arXiv:1603.02233 [gr-qc]].

\bibitem{Cho:2003qe}
H.~T.~Cho,
Phys. Rev. D \textbf{68}, 024003 (2003)
doi:10.1103/PhysRevD.68.024003
[arXiv:gr-qc/0303078 [gr-qc]].

\bibitem{Turimov:2020fme}
B.~Turimov, J.~Rayimbaev, A.~Abdujabbarov, B.~Ahmedov and Z.~Stuchl\'\i{}k,
Phys. Rev. D \textbf{102}, no.6, 064052 (2020)
doi:10.1103/PhysRevD.102.064052
[arXiv:2008.08613 [gr-qc]].


\bibitem{Dariescu:2023twk}
M.~A.~Dariescu, V.~Lungu, C.~Dariescu and C.~Stelea,
Phys. Rev. D \textbf{109}, no.2, 024021 (2024)
doi:10.1103/PhysRevD.109.024021
[arXiv:2311.11356 [gr-qc]].

\bibitem{Lim:2021ejg}
Y.~K.~Lim,
Phys. Rev. D \textbf{103}, no.8, 084044 (2021)
doi:10.1103/PhysRevD.103.084044
[arXiv:2102.08531 [gr-qc]].

\bibitem{Lim:2022qrt}
Y.~K.~Lim,
Phys. Rev. D \textbf{106}, no.6, 064023 (2022)
doi:10.1103/PhysRevD.106.064023
[arXiv:2206.00170 [gr-qc]].

\bibitem{Dariescu:2010zz}
M.~A.~Dariescu and C.~Dariescu,
Astropart. Phys. \textbf{34}, 116-120 (2010)
doi:10.1016/j.astropartphys.2010.06.005


\bibitem{Richarte:2021fbi}
M.~G.~Richarte, \'E.~L.~Martins and J.~C.~Fabris,
Phys. Rev. D \textbf{105}, no.6, 064043 (2022)
doi:10.1103/PhysRevD.105.064043
[arXiv:2111.01595 [gr-qc]].


\bibitem{Birkandan:2017ypz}
T.~Birkandan and M.~Horta\c{c}su,
Gen. Rel. Grav. \textbf{50}, no.3, 28 (2018)
[erratum: Gen. Rel. Grav. \textbf{50}, no.9, 109 (2018)]
doi:10.1007/s10714-018-2351-y
[arXiv:1711.06811 [gr-qc]].

\bibitem{Birkandan:2020ivm}
T.~Birkandan and M.~Hortacsu,
Eur. Phys. J. C \textbf{81}, no.5, 389 (2021)
doi:10.1140/epjc/s10052-021-09182-w
[arXiv:2012.07352 [gr-qc]].

\bibitem{Vieira:2018djw}
H.~S.~Vieira, V.~B.~Bezerra, C.~R.~Muniz and M.~S.~Cunha,
Int. J. Mod. Phys. D \textbf{28}, no.12, 1950151 (2019)
doi:10.1142/S0218271819501517
[arXiv:1807.09135 [gr-qc]].

\bibitem{Dariescu:2018rrr}
M.~A.~Dariescu, C.~Dariescu and C.~Stelea,
Adv. High Energy Phys. \textbf{2019}, 5769564 (2019)
doi:10.1155/2019/5769564
[arXiv:1812.06852 [hep-th]].

\bibitem{Stelea:2011fj}
C.~Stelea, K.~Schleich and D.~Witt,
Phys. Rev. D \textbf{91}, 024040 (2015)
doi:10.1103/PhysRevD.91.024040
[arXiv:1108.5145 [gr-qc]].

\end{thebibliography}
\end{document}